\documentclass[a4paper,fleqn,usenatbib, singlespacing, twocolumn]{mnras}

\usepackage{newtxtext,newtxmath}

\usepackage[T1]{fontenc}
\usepackage{ae,aecompl}
\usepackage{threeparttable}

\usepackage{graphicx}	
\usepackage{amsmath}	
\usepackage{amssymb}	
\usepackage{bm}

\title[Radio emission from ultracool dwarfs]{Auroral Radio Emission from Ultracool Dwarfs: a Jovian Model}

\author[S. Turnpenney et al.]{
S. Turnpenney,$^{1}$\thanks{E-mail: st349@le.ac.uk}
J. D. Nichols,$^{1}$
G. A. Wynn,$^{1}$
S. L. Casewell$^{1}$
\\
$^{1}$Department of Physics and Astronomy, University of Leicester, Leicester, LE1 7RH, UK\\
}

\date{Accepted XXX. Received YYY; in original form ZZZ}

\pubyear{2016}

\begin{document}
\label{firstpage}
\pagerange{\pageref{firstpage}--\pageref{lastpage}}
\maketitle

\begin{abstract}
A number of fast-rotating ultra cool dwarfs (UCDs) emit pulsed coherent radiation, attributed to the electron cyclotron maser instability, a phenomenon that occurs in the solar system at planets with strong auroral emission.  In this paper we examine magnetosphere-ionosphere coupling currents in UCDs, adopting  processes used in models of Jovian emission.  We consider the angular velocity gradient arising from a steady outward flux of angular momentum from an internal plasma source, as analogous to the jovian main oval current system, as well as the interaction of a rotating magnetosphere with the external medium. Both of these mechanisms are seen in the solar system to be responsible for the production of radio emission.  We present the results of an investigation over a range of relevant plasma and magnetosphere-ionosphere coupling parameters to determine regimes consistent with observed UCD radio luminosities. Both processes are able to explain observed UCD luminosities with ionospheric Pedersen conductances of $\sim 1-2$ mho, either for a closed magnetosphere with a plasma mass outflow rate of $\sim 10^5 \, \mathrm{kg}$ s$^{-1}$, i.e. a factor of $\sim 100$ larger than that observed at Jupiter's moon Io, or for a dwarf with an open magnetosphere moving through the interstellar medium at $\sim 50 \, \mathrm{km}$ s$^{-1}$ and a plasma mass outflow rate of $\sim 1000 \, \mathrm{kg}$ s$^{-1}$.  The radio luminosity resulting from these mechanisms have opposing dependencies on the magnetic field strength, a point which may be used to discriminate between the two models as more data become available.
\end{abstract}

\begin{keywords}
brown dwarfs -- stars: late-type -- stars: low-mass -- stars: magnetic field -- plasmas -- radio continuum: stars.
\end{keywords}



\section{Introduction} \label{sec:intro}

Ultracool dwarfs (UCDs) are objects with spectral type M7 and later, a class encompassing brown dwarfs along with the lowest mass stars \citep{kirkpatrick1997}. With radii similar to Jupiter, and masses of $\sim$13 - 90 M$_\mathrm{Jup}$, UCDs bridge the stellar and planetary regimes.  Observations have revealed intense radio emission in the GHz frequency range in $\sim$ 10\% of the UCDs targeted in radio surveys \citep{berger2001, berger2002, berger2009, hallinan2006, hallinan2008, phan-bao2007, antonova2008, antonova2013, lynch2015, lynch2016, burningham2016}. Quiescent and flaring radio emission was first reported by \citet{berger2001}, with a luminosity approximately four orders of magnitude greater than predicted by the G\"udel-Benz relation \citep{guedel1993}, an empirical relation between the observed X-ray and radio luminosities of stellar objects over a wide range of spectral classes.  This implies that the cool, relatively neutral atmospheres of these UCDs do not preclude significant magnetic activity. Coherent pulsed radio emission was later detected at an M9 UCD, TVLM 513-46546 (hereafter TVLM-513), modulated at the rotation period of the dwarf \citep{hallinan2006}.  Subsequent observations have revealed pulsed radio activity present in other M, L and T dwarfs \citep{berger2009, route2012, route2016, williams2015, kao2016}. This pulsed emission is atypical of stellar radio emission in that it was found to be almost completely circularly polarized, with a high brightness temperature, and highly beamed, and was attributed to the electron-cyclotron maser instability (ECMI) as the generation mechanism \citep{hallinan2008}. In every case where coherent pulsed emission has since been detected at UCDs, quiescent emission is also present. It has been suggested that this may be an unpolarised component of ECMI emission \citep{hallinan2008}, although gyrosynchrotron emission is also a candidate mechanism \citep{berger2002}. In situ measurements have found ECMI emission present at solar system planets: at Earth in the auroral kilometric radiation \citep{wu1979, ergun2000, treumann2006}; decametric radiation at Jupiter \citep{imai2008}; and Saturnian kilometric radiation \citep{lamy2011}. It has also been suggested that ECMI emission may be present at exoplanets \citep[e.g.][]{zarka2007, griessmeier2007, nichols2011, nichols2016}.

Radio emission provides a means of diagnosing properties of the magnetic fields at UCDs, and in the case of later spectral type objects, where measurements of Zeeman broadening become ineffective, radio emission is an important diagnostic of the magnetic field strength and topology. The presence of $\sim$GHz ECMI emission implies at least kilogauss magnetic fields at those UCDs observed to emit at these frequencies. In solar-type stars the magnetic field is generated by a dynamo mechanism operating at the radiative-convective boundary \citep{parker1955}. As UCDs are fully convective, this mechanism is precluded and other dynamo models are suggested \citep{dobler2006, browning2008}. 
  
Potential source models for UCD ECMI-induced emission include modelling active sectors containing coronal loops along which hot plasma precipitates \citep{yu2011, lynch2015}, and interaction of the dwarf with a satellite \citep{kuznetsov2012}. Numerical modelling of satellite-induced ECMI is inconsistent with observations, since \citet{kuznetsov2012} show that it predicts variable time intervals, rather than the observed regular intervals between pulses.  Modelling the emission from an active sector containing coronal loops is able to simulate the observed emission, but these phenomena would likely be transient, whereas the coherent pulsed emission has been found to be steady and persistent. Additionally, the coronal loop model does not directly account for the observed increase in radio activity with faster rotation \citep{mclean2012} and it is known that ECM generation is more efficient for electrons that are accelerated by magnetic field-aligned electric potentials \citep{treumann2006}.   

Recently, \citet{hallinan2015} discovered optical variation of the dwarf LSR J1835+3259 in phase with radio emissions and attributed these effects to auroral activity.  This suggests the existence of a system of currents that give rise to these emissions. At Jupiter and other magnetised planets in the solar system, field-perpendicular currents in the Pedersen layer of the atmosphere (where the gyrofrequency is approximately equal to the collision frequency) arise due to the motion of plasma relative to the neutral particles, i.e. from $\mathbf{E = - v \times B}$. If there is a gradient in the angular velocity of the plasma then a divergence of the Pedersen current occurs, resulting in a field-aligned current system coupling the ionosphere to the magnetosphere. The gradient in the plasma angular velocity at Jupiter is due to centrifugally driven outward flow of plasma originating from a torus of material volcanically liberated from the moon Io, orbiting deep within the planet's magnetosphere \citep{hill1979, cb2001, cnb2002, nichols2003, nichols2004, nichols2005}. Such field-aligned currents have been shown to be associated with with the generation of ECMI emission \citep{treumann2006} and are the process driving the emission from Jupiter's main auroral oval.  It may therefore be reasonably inferred from the observed ECMI emission that equivalent electric current systems flow in the magnetospheres of some UCDs.  \citet{schrijver2009} hypothesised that ECMI emission at UCDs could arise from a current system similar to that which generates Jupiter's main auroral oval.  \citet{nichols2012} (hereafter N12) modelled the radio emission as a manifestation of downward-precipitating electrons on upward-flowing field-aligned currents, and demonstrated that a Jupiter-like magnetosphere-ionosphere (M-I) coupling current system can broadly explain the properties of the radio emission at UCDs targeted in their study. However, N12 considered only a representative angular velocity gradient in the ionospheric plasma, conveniently denoted using a tanh function, and did not explicitly compute the angular velocity profiles expected at UCDs with a Jupiter-like current system.

In this paper we model two separate processes for the generation of ECMI-induced auroral emission at UCDs.  Firstly, we develop the model of N12 by calculating the angular velocity profile resulting from a steady outward flow of plasma mass from an internal source within a closed magnetosphere.  At Jupiter, this replenishing plasma source is provided by the moon Io and a similar plasma source from a close-in companion is conceivable at UCDs.  We also use a second model, in which a flow shear occurs at an open-closed field line boundary (OCFB). Since the models represent contrasting processes that give rise to auroral emission, the results for both are examined over a wide range of key system parameters in order to constrain the physical origin of the observed radio emissions.

\section{Theoretical analysis}

In this section we outline the theoretical basis for the UCD model used in this study.  The analysis comprises four elements: a model of the precipitating electron acceleration necessary to drive the auroral radio emission; a description of the magnetic field model employed; calculation of the M-I coupling current system; and determination of the angular velocity of the magnetospheric plasma.

\subsection{Auroral emission}

Auroral emissions are driven by the precipitation of energetic electrons down magnetic field lines and onto the atmosphere. The total power $P_e$ from this process can be determined by integrating the precipitating electron energy flux $E_f$ over one hemisphere of the UCD, i.e.
\begin{equation} \label{eq21}
P_e = \int^{90}_0 2 \pi R^2_{\mathrm{UCD}} \sin \theta_i E_f d \theta_i,
\end{equation}
where $R_{\mathrm{UCD}}$ is the radius of the dwarf and $\theta_i$ is the ionospheric colatitude in degrees. If we assume that the radiation is visible from only one hemisphere during an observation, and that, in common with observations at Jupiter and Saturn, there is a 1\% generation efficiency of the electron-cyclotron maser instability \citep{gustin2004, clarke2009, lamy2011}, we can calculate the spectral luminosity, $L_r$:

\begin{equation} \label{eq22}
L_r = \frac{P_e}{100 \Delta \nu},
\end{equation}
where $\Delta \nu$ is the bandwidth, assumed to be equal to the electron-cyclotron frequency in the polar ionosphere and hence given by
\begin{equation}
\Delta \nu = \frac{e B_i}{2 \pi m_e}.
\end{equation}

Qualitatively, the precipitating electron energy flux $E_f$ is the kinetic energy flux of downward flowing electrons carrying a field-aligned current.  In general these field-aligned currents, which will be described in Section \ref{FAC}, cannot be carried by unaccelerated precipitating electrons alone.  There is a maximum current density that can carried by unaccelerated electrons, which for an isotropic Maxwellian distribution in velocity space is given by 

\begin{equation} \label{eq17}
j_{\|i0} = e n \left(\frac{W_{th}}{2 \pi m_e} \right)^{1/2},
\end{equation}
where $e$ and $m_e$ are the charge and mass of the electron respectively, and $W_{th}$ and $n$ are the thermal energy and number density of the electron source population respectively \citep{lundin1978}. This arises when all downward precipitating electrons are absorbed by the atmosphere, resulting in an empty upward-going hemisphere in velocity space, and a full downward-going hemisphere. In general this current is very low in all planetary magnetospheres. Hence, a field-aligned current density which exceeds the value given by equation (\ref{eq17}) requires a field-aligned accelerating potential, the magnitude of which is determined using \citeauthor{knight1973}'s theory (\citeyear{knight1973}) of parallel electric fields, developed originally for application to Earth but has since been applied to Jupiter and Saturn, where required voltages are found to be consistent with the energies of precipitating electrons. Since the results shown in this paper necessitate accelerating potentials significantly more powerful than those found at solar system planets, comparable to and exceeding the rest mass of the electron ($\sim$ 511 keV), we employ \citeauthor{cowley2006}'s (\citeyear{cowley2006}) relativistic extension of Knight's theory, for which the relation between voltage and field-aligned current density is given by 

\begin{equation} \label{eq18}
\left( \frac{j_{\|i}}{j_{\|i0}} \right) = 1 + \left(\frac{e \Phi_{\mathrm{{min}}}}{W_{th}}  \right) + \frac{\left(\frac{(e \Phi_{\mathrm{min}}}{W_{th}}\right)^2}{2 \left[ \left(\frac{m_e c^2}{W_{th}} \right)+ 1\right]},
\end{equation}
 where $c$ is the speed of light in a vacuum and $\Phi_{\mathrm{min}}$ is the minimum field-aligned potential required to drive the current $j_{\| i}$. The minimum potential requires that the magnetic field strength at the accelerator is much less than the field strength in the ionosphere, a condition easily satisfied owing to the strong $r^{-3}$ dependence of a dipole field strength on radial distance. For $j_{\| i} < j_{\| i 0}$ we have $\Phi_{\mathrm{min}} = 0$, since an upward current with $j_{\| i} < j_{\| i 0}$ can be carried by unaccelerated magnetospheric electrons as discussed above, and we assume that downward current can be easily carried by abundant ionospheric electrons and thus does not yield auroral radio emissions. The corresponding precipitating electron energy flux carried by the field-aligned current computed in equation (\ref{eq18}) was derived by \citet{cowley2006}, and here we employ the result from that work:
\begin{equation}
\frac{E_f}{E_{f0}} = 1 +\left( \frac{e\Phi_{\mathrm{min}}}{W_{th}} \right) + \frac{1}{2}  \left( \frac{e\Phi_{\mathrm{min}}}{W_{th}} \right)^2 + \frac{ \left( \frac{e\Phi_{\mathrm{min}}}{W_{th}}  \right)^3 }{ 2 \left[ 2 \left( \frac{m_e c^2}{W_{th}} \right) + 3 \right]},
\end{equation}
where $E_{f0}$ is the maximum energy flux of the precipitating electrons in an unaccelerated population, corresponding to equation (\ref{eq17}), i.e.

\begin{equation}
E_{f0} = 2 n W_{th} \left( \frac{W_{th}}{2 \pi m_e} \right)^{1/2}.
\end{equation}

\subsection{Magnetic Field Model}
 
Turning now to the magnetic field model used in this study, in common with N12 and previous work studying Jupiter's rotationally dominated middle magnetosphere \citep[e.g.][]{cb2001, cnb2002}, we assume an initially azimuthally symmetric, spin-aligned magnetic field. Note that we expect this field polarity in 50\% of dwarfs, and we go on to consider the effect of opposite polarity below.   The magnetic field components can then be specified by a flux function $F(\rho, z)$, related to the field components by $\mathbf{B} = (1/ \rho) \nabla{F} \times \bm{\hat{\varphi}}$, where $z$ is the distance along the magnetic axis from the equator,  $\rho$ is the perpendicular distance from this axis, and $\bm{\hat{\varphi}}$ is the azimuthal angle.  A flux shell of field lines flowing from the northern to southern ionosphere via the equator is defined by $F$=constant, such that mapping of the field lines between the ionosphere and the equatorial plane is achieved simply by writing $F_e = F_i$, where the subscripts denote equatorial and ionospheric flux functions respectively. We assume that the ionosphere is overwhelmingly dominated by the planetary dipole, for which the flux function is

\begin{equation} \label{eq1}
F_{dip} = B_{\mathrm{UCD}} \; \rho^2 \; \left( \frac{R_{\mathrm{UCD}}}{r} \right)^3,
\end{equation}
where $B_{\mathrm{UCD}}$ is the equatorial surface magnetic field strength of the UCD in Tesla (note here and throughout we use SI units), $r$ is the radial distance from the centre of the dwarf. Here we have assumed $F = 0$ on the dipole axis. The flux function in the polar ionosphere is therefore
\begin{equation} \label{eq3}
F_i = B_{\mathrm{UCD}} \; \rho_i^2 = B_{\mathrm{UCD}} \; R_{\mathrm{UCD}}^2 \; \sin^2 \theta_i,
\end{equation}
where $\rho_i$ is the perpendicular distance from the dipole axis.  Equations (\ref{eq1}) and (\ref{eq3}) allow convenient mapping between the magnetosphere and ionosphere, such that a field line threading the equatorial plane maps to an ionospheric colatitude given by 

\begin{equation} \label{eq5a}
\sin \theta_i = \sqrt{\frac{F_e (\rho_e)}{B_{\mathrm{UCD}} \; R^2_{\mathrm{UCD}}}}.
\end{equation}\\ \\
In general, the field at larger distances is modified by magnetospheric currents. For example magnetopause currents act to confine the field to a finite volume.  The radius of the magnetopause is governed by pressure balance, and we can estimate the size of the magnetosphere using 

\begin{equation} \label{eq24}
\left( \frac{R_{\text{mp}}}{R_{\text{UCD}}} \right) = \left( \frac{k^2_m B_{\text{UCD}}^2}{2 \mu_{\circ} (p_{th} + p_{\text{dyn}} + p_B)} \right)^{1/6},
\end{equation}
where $p_{th}$,  $p_{dyn}$ and $p_B$ are respectively the thermal, dynamic and magnetic pressure of the local interstellar cloud outside the magnetosphere, and $k_m$ is factor by which magnetopause currents enhance the magnetic field and is dependent on the geometry at the boundary. Since the dipole extends to infinity, we modify the flux function, and hence the field component $B_{ze}$, with an additional term allowing us to simulate closed and open magnetic fields. The flux function in the equatorial plane is thus

\begin{equation} \label{eq26}
F_e = \frac{B_{\mathrm{UCD}} R_{\mathrm{UCD}}^3}{\rho_e} - \frac{(k_m - 1) B_{\mathrm{UCD}} \rho_e^2 R_{\mathrm{UCD}}^3}{2 R_{\mathrm{mp}}^3} + F_{e \infty},
\end{equation}
and the magnitude of the north-south field component threading the equatorial plane is given by

\begin{equation}
|B_{ze}| = B_{\mathrm{UCD}} \left( \frac{R_{\mathrm{UCD}}}{\rho_e} \right)^3 + \frac{(k_m - 1) B_{\mathrm{UCD}}^2 R_{\mathrm{UCD}}^3}{R_{\mathrm{mp}}^3},
\end{equation} 
where $F_{e \infty}$ is the equatorial flux function at infinity.   To model a closed magnetosphere with an internal dipole we set $k_m$ = 3, which is the value required to confine a dipole within a sphere, and is such that the compressed field at the magnetopause is exactly three times the uncompressed value \citep{parker1962}. We also set $F_{e \infty}$ = 0, such that the field line at the magnetopause maps to the dipole axis.  Alternatively, modelling the scenario of an open magnetosphere, in which interaction of the magnetopause with the interstellar medium occurs by the process of magnetic reconnection, is achieved simply by using equation (\ref{eq5a}) to set the value of $F_{e \infty}$ according to the colatitude of the OCFB.

\subsection{Magnetosphere-ionosphere coupling current system} \label{FAC}

\begin{figure}
	\includegraphics[width=\columnwidth, trim = 0 0 0 0]{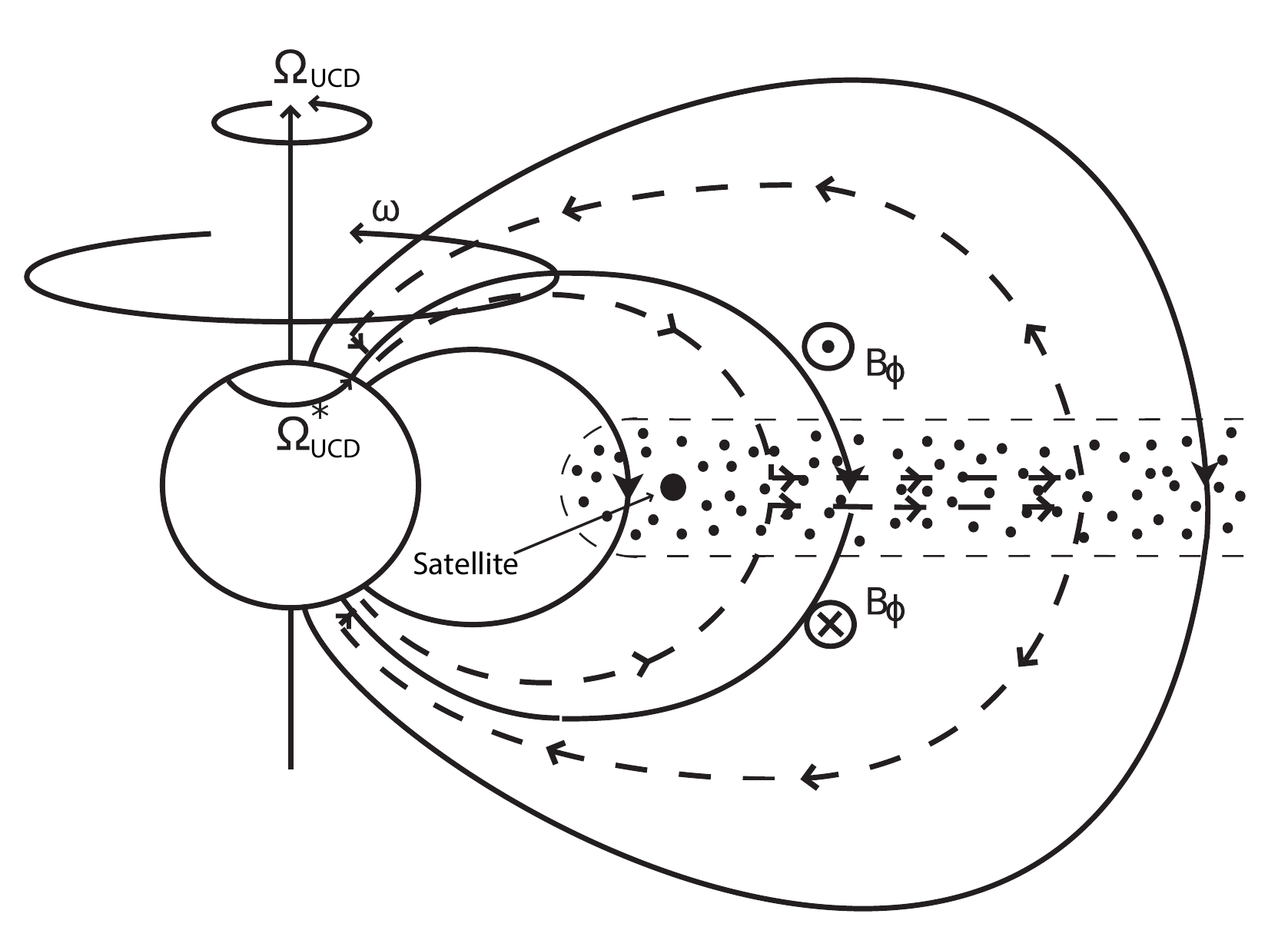}
    \caption{A schematic of the closed magnetosphere used in this model viewed in a meridian cross-section.  The radially outward flowing plasma originating from a satellite orbiting the UCD is indicated by the dotted region. The solid black lines represent the compressed closed magnetic field.  The arrowed dashed lines indicate the current system flowing between the ionosphere and magnetosphere.  The azimuthal magnetic field components arising due to a bending of the field lines by the M-I current system are indicated flowing into and out of the diagram. The outer boundary of the magnetosphere is compressed due to currents flowing at the magnetopause. Modified from Figure 1 of \citet{cb2001}}
    \label{fig:schematic}
\end{figure}

The current system coupling the magnetosphere and the ionosphere and communicating torque between the two is illustrated by the dashed lines in Fig. \ref{fig:schematic}, for the jovian system.  The formulation employed in calculating these currents has been extensively applied at Jupiter \citep{hill1979, pontius1997, cb2001, cnb2002, nichols2003}.  Here we provide expressions for the relevant currents, deferring the full details of the formulation to Appendix A.

For an equatorial plasma of angular velocity $\omega$, the total current flowing in the Pedersen layer of the atmosphere is 

\begin{equation}
I_P =  4 \pi \Sigma_P^* \Omega_{\mathrm{UCD}} F_e \left( 1 - \frac{\omega}{\Omega_{\mathrm{UCD}}} \right),
\end{equation}
where $\Sigma_P^*$ is the `effective' Pedersen conductivity, and $\Omega_{\mathrm{UCD}}$ is the angular velocity of the dwarf. Current continuity and the north-south symmetry between the two hemispheres shown in Fig. \ref{fig:schematic} imply that the total radial current $I_{\rho}$ is equal to $2 I_P$. The field-aligned current density follows directly from the divergence of either the Pedersen or radial current, and has a value just above the ionosphere of 

\begin{equation} \label{eq12}
j_{\| i} = - \frac{4 B_{\mathrm{UCD}} \Omega_{\text{UCD}}}{\rho_e B_{ze}}   \frac{d}{d\rho_e} \left[ \Sigma_p^* \left( 1 - \frac{\omega}{\Omega_{\text{UCD}}} \right) F_e \right].
\end{equation}
In the absence of models of the ionospheres of UCDs, we assume that the Pedersen conductance is a constant, although this would be expected to vary because of modulation by auroral precipitation. We will address this point in future work.

\subsection{Angular velocity profile}

Our analysis requires the determination of the steady state angular velocity profile of the rotating plasma. In previous work modelling auroral currents at UCDs \citep{nichols2012} the angular velocity profile was a simple representative function, here we calculate the ionospheric angular velocity profile following a formulation originally proposed by \cite{hill1979} for the jovian system. This formulation  assumes a steady source of plasma within the magnetosphere of the dwarf, an assumption which will be examined later in this paper.  As the newly-created plasma is picked up by the corotating magnetic field it becomes centrifugally unstable and diffuses radially outward via flux tube interchange.  Conservation of angular momentum would cause the angular velocity of the plasma to fall as $r^{-2}$ if no torques acted.  However, the current system considered in the previous section transfers angular momentum from the dwarf to the magnetosphere, enforcing a state of partial corotation.  Following \citet{hill1979, hill2001}, \cite{pontius1997}, \cite{nichols2003, nichols2004, nichols2005} and \cite{cnb2002} application of Newton's second law to a steady diffuse outflow of plasma from the vicinity of the dwarf yields

\begin{equation} \label{eq15}
\frac{d}{d \rho_e}(\rho^2_e  \omega(\rho_e))= \frac{2 \pi \rho^2_e i_\rho |B_{ze}|}{\dot{M}},
\end{equation}
where $\dot{M}$ is the plasma mass outflow rate, assumed to be constant. The left-hand side of equation (\ref{eq15}) is the radial gradient of the plasma angular momentum per unit mass, and the right-hand side is the ionospheric torque on the equatorial plasma.  Substituting equation (\ref{eq9}) and expanding gives

\begin{equation} \label{eq14}
\frac{\rho_e}{2} \frac{d}{d \rho_e} \left( \frac{\omega}{\Omega_{\mathrm{UCD}}} \right) + \left( \frac{\omega}{\Omega_{\mathrm{UCD}}} \right) = \frac{4 \pi \Sigma_p^* F_e |B_{ze}|}{\dot{M}} \left(1- \frac{\omega}{\Omega_{\mathrm{UCD}}} \right),
\end{equation}
which we refer to as the `Hill-Pontius' (H-P) equation. This equation must generally be solved numerically with the use of one boundary condition. The `Hill distance' (not to be confused with the radius of the `Hill sphere' within which a body's gravitational field dominates) is the characteristic distance over which the angular velocity of the plasma departs from rigid corotation, and for a dipole field is given by

\begin{equation} \label{eq16}
\frac{R_H}{R_{\text{UCD}}} =\left( \frac{2 \pi \Sigma_P^* B^2_{\mathrm{UCD}} R^2_{\mathrm{UCD}}}{\dot{M}}\right)^{1/4}.
\end{equation}
For the modified dipole fields under consideration here, equation (\ref{eq16}) serves as reasonable zeroth-order approximation.

Clearly the form of the angular velocity profile, and consequently the field-aligned current density, is governed by the M-I coupling parameters, $\Sigma_P^*$ and $\dot{M}$, and the effect of varying the values of these parameters will be examined in the results and discussion below. 

We further consider an open magnetosphere, containing a substantial quantity of open flux. Fig. \ref{fig:isbell} taken from \citet{isbell1984} shows a schematic of the the helical open magnetic field configuration resulting from an external flow past a rotating object. Such a system is analogous to a Faraday disc, and \citet{isbell1984} showed that the angular velocity of the open field lines is:

\begin{figure}
	\includegraphics[width=\columnwidth, trim = 0 0 0 0]{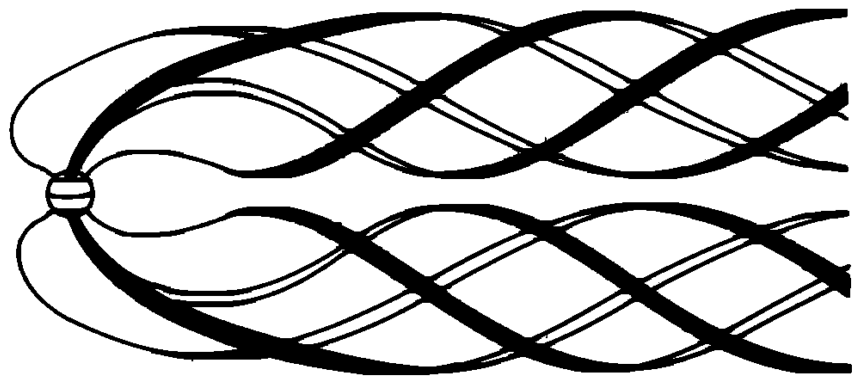}
    \caption{Sketch illustrating the helical configuration of the open magnetic field lines arising from an external flow over a rotating magnetised object.  From \citet{isbell1984}}
    \label{fig:isbell}
\end{figure}

\begin{equation} \label{eqIsbell}
\left( \frac{\omega}{\Omega_{\mathrm{UCD}}} \right)_{\mathrm{open}} = \frac{\mu_0 \Sigma_P^* V_{\mathrm{ISM}}}{1 + \mu_0 \Sigma_P^* V_{\mathrm{ISM}}},
\end{equation}
where $\mu_0$ is the permeability of free space, and $V_{\mathrm{ISM}}$ is the velocity of the interstellar medium relative to the dwarf. . To transition from the angular velocity of the closed field lines to that of the open field lines we follow previous work \citep{cowley2005} and employ a representative tanh function, given by

\begin{equation} \label{eqtanh}
\left( \frac{\omega}{\Omega_{\mathrm{UCD}}} \right) = \left( \frac{\omega}{\Omega_{\mathrm{UCD}}} \right)_{\mathrm{open}} + \frac{\Delta \omega}{2} \times \left[ 1 + \mathrm{tanh} \left( \frac{\theta_i - \theta_{ic}}{\Delta \theta_i} \right) \right],
\end{equation}
where $(\omega / \Omega_{\mathrm{UCD}})_{\mathrm{open}}$ is the angular velocity of the open magnetic field lines, given by equation (\ref{eqIsbell}), $\theta_{ic}$ is the colatitude of the centre of the transition region, $\Delta \theta_i$ is the half-width of the of the transition region, and $\Delta \omega$ is the difference between the high and low values between which the angular velocity transitions.  The solution to the Hill-Pontius equation is joined to the profile given by equation (\ref{eqtanh}) by ensuring that $\omega$ and $\frac{d\omega}{d \theta_i}$ are continuous at the OCFB. The role of the plasma mass outflow rate in the open model is to reduce the angular velocity of the equatorial plasma in the closed magnetosphere, as in the closed magnetosphere model. In both models the plasma is eventually lost via the pinching off of plasmoids at low latitudes, and is not evisaged to modify the dynamics of the higher latitude open field lines. The location of the OCFB is independent of $\dot{M}$, and instead is related to the amount of open magnetic flux in the system.  Empirically, the location of the OCFB at Earth, Saturn, and probably Jupiter also, is $\sim$ 10$^{\circ}$ colatitude, and therefore we have assumed the same value in this work.  We note that we envisage plasma mass inflow from the ISM to be negligible in the open magnetosphere, which is dominated by outward-flowing plasma of internal origin.  The fraction of incident particles absorbed by the Earth's magnetosphere via reconnection is $10^{-3}$ \citep{hill1983}. The high latitude open field lines will thus contain plasma of mixed UCD and ISM origins, though by analogy with planetary magnetospheres we expect the density on open field lines to be low.  As stated above, the angular velocity of the open field region is dominated by the conductance of the ionosphere and velocity of the UCD relative to the ISM.

\section{Results}

\subsection{Representative values} \label{sec:representative}

Although the plasma populations at UCDs are unknown, we can examine a range of values for the plasma and M-I coupling parameters, observing how the angular velocity profile, current system, and radio emission vary in response, in order to estimate the likely parameters at observed UCDs. 

\begin{figure}
	\includegraphics[width=\columnwidth, trim = 0 0 0 100]{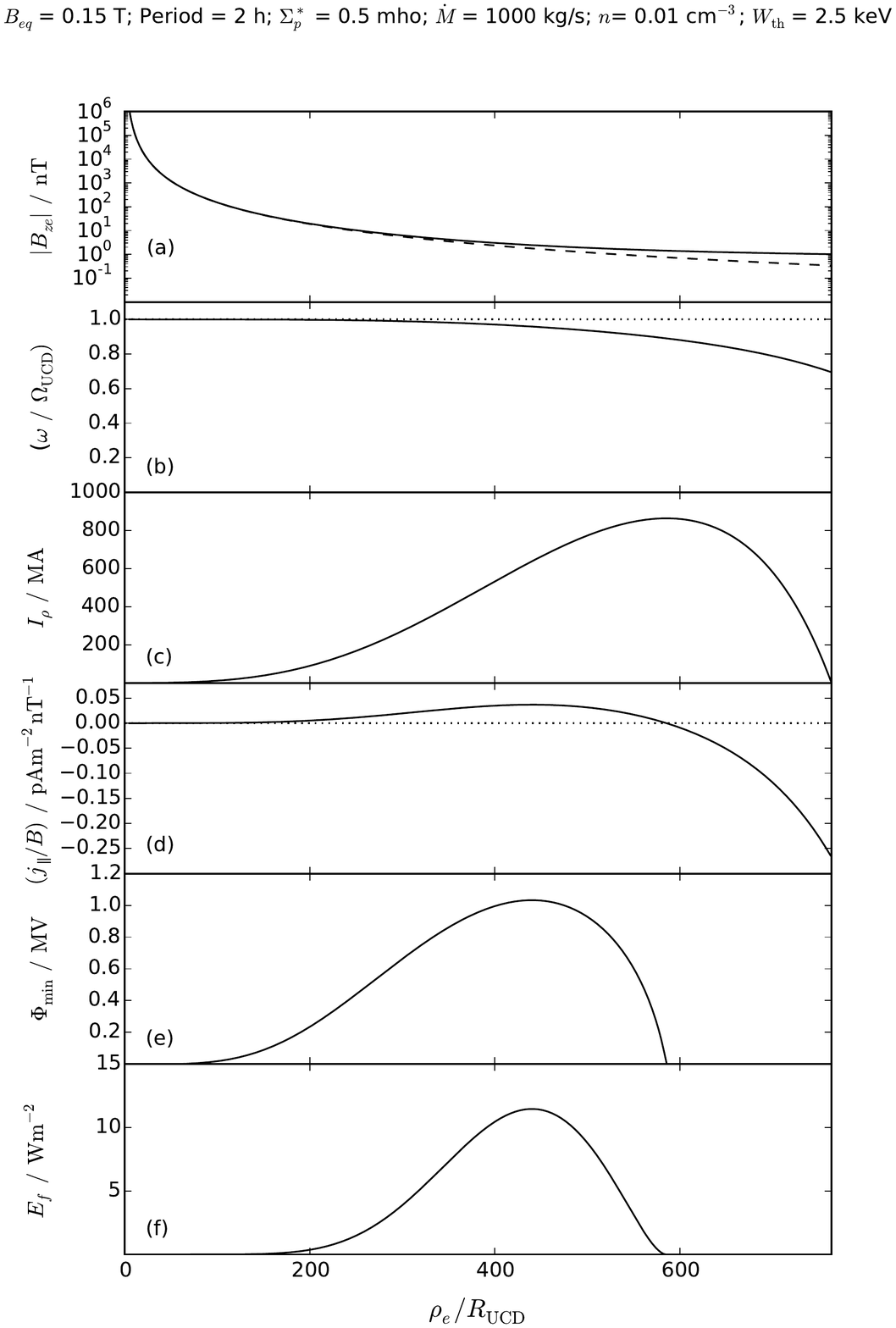}
    \caption{Profiles of the system parameters for the closed magnetic field model featured in this paper, shown over an equatorial radial range corresponding to a magnetic field strength of 0.15 T: (a) the $z$-component of the field strength in the equatorial plane, with the dipole field represented by the dashed line, (b) the angular velocity $(\omega / \Omega_{\text{UCD}})$, (c) the radial current flowing in the equatorial plane in GA, (d) the field-aligned current density per unit magnetic field strength $(j_\| / B)$, (e) the minimum field-aligned electric potential in MV required to accelerate electrons and drive the current just above the ionosphere, and (f) the precipitating electron energy flux $E_f$ in Wm$^{-2}$.  The values of the magnetic field strength, rotation period, M-I coupling and plasma parameters used are indicated at the top of the figure.}
    \label{fig:equatorial}
\end{figure}

Before investigating a range of values in parameter space, we first demonstrate results of the model for both an open and closed magnetosphere using representative jovian values for the M-I coupling and plasma parameters. Fig. \ref{fig:equatorial} shows the M-I coupling parameters versus radial distance $\rho_e$ for a closed magnetic field with an equatorial field strength of 0.15 T (equal to 1.5 kG) and rotation period of $\sim$ 2 h, consistent with values observed for TVLM-513 \citep{hallinan2008}. The radial limit of the model is set at 713 R$_{\mathrm{UCD}}$, a value determined from pressure balance between the magnetospheric magnetic pressure and the interstellar medium parameters as discussed above. Here and below we have taken typical values of 0.5 nT for the magnetic field strength, $7 \times 10^4 \, \mathrm{m}^{-3}$ for the number density and $7 \times 10^3$ K for the temperature of the plasma in the local interstellar cloud, and assumed the UCD has a velocity of 50 km s$^{-1}$ with respect to the medium \citep{vanhamaki2011}.  Fig. \ref{fig:equatorial}(a) shows the magnitude of the north-south component of the magnetic field in the equatorial plane $|B_{ze}|$, which is larger than for the dipole and falls to a value of $\sim $ 1 nT at the outer boundary. Fig. \ref{fig:equatorial}(b) shows the plasma angular velocity $(\omega / \Omega_{\mathrm{UCD}})$ resulting from a numerical solution of equation (\ref{eq14}), where we have taken spot values of $\dot{M} = 10^3 \;\mathrm{kg/s} $ and $\Sigma_p^* = 0.5$ mho. The dipole Hill distance given by equation (\ref{eq16}) for these M-I coupling parameters is 775 R$_{\mathrm{UCD}}$, therefore lying outside the magnetosphere. The plasma near-rigidly corotates with the dwarf throughout most of the magnetosphere, only falling to an angular velocity of $\sim$0.7 $\Omega_{\text{UCD}}$ at the magnetopause. In this case, therefore, the angular velocity shear is significantly less than in the fiducial model of N12.   Fig. \ref{fig:equatorial}(c) shows the corresponding equatorial radial current $I_{\rho}$ that flows as a result of the departure from corotation.  The current rises with radial distance to a peak value of 865 MA  at 586 $\mathrm{R_{UCD}}$ before returning to zero at the magnetopause due the closing of the current system within the magnetosphere.  The field-aligned current density per unit magnetic field strength $(j_{\|} / B)$, a quantity that is constant along a field line, is shown in Fig. \ref{fig:equatorial}(d)  and is found to rise to positive peak, i.e. upward-flowing current associated with auroral emission, of $\sim$0.037 pAm$^{-2}$ nT$^{-1}$ at 440 $\mathrm{R_{UCD}}$ before transitioning to large negative values in the outer region of the magnetosphere, peaking at -0.266 pAm$^{-2}$ nT$^{-1}$ at the magnetopause. Fig. \ref{fig:equatorial}(e) shows the minimum field-aligned electric potential required to drive the upward-flowing field-aligned current, $\Phi_{\mathrm{min}}$, as given by equation (\ref{eq18}), and where we have employed plasma number density $n$ and thermal energy $W_{th}$ values of 0.1 cm$^{-3}$ and 2.5 keV respectively, but examine wider ranges of these parameters below. The field-aligned voltage peaks at 1.03 MV at the same radial distance as the peak in the current density, and drops to zero beyond 580 R$_{\mathrm{UCD}}$  Finally, the precipitating electron energy flux $E_f$ is shown in Fig. \ref{fig:equatorial}(f), peaks at $ 11.4 \; \mathrm{ W m}^{-2}$, the same radial distance as the peak in field-aligned current density and voltage. 

\begin{figure}
	\includegraphics[width=\columnwidth, trim = 0 0 0 100]{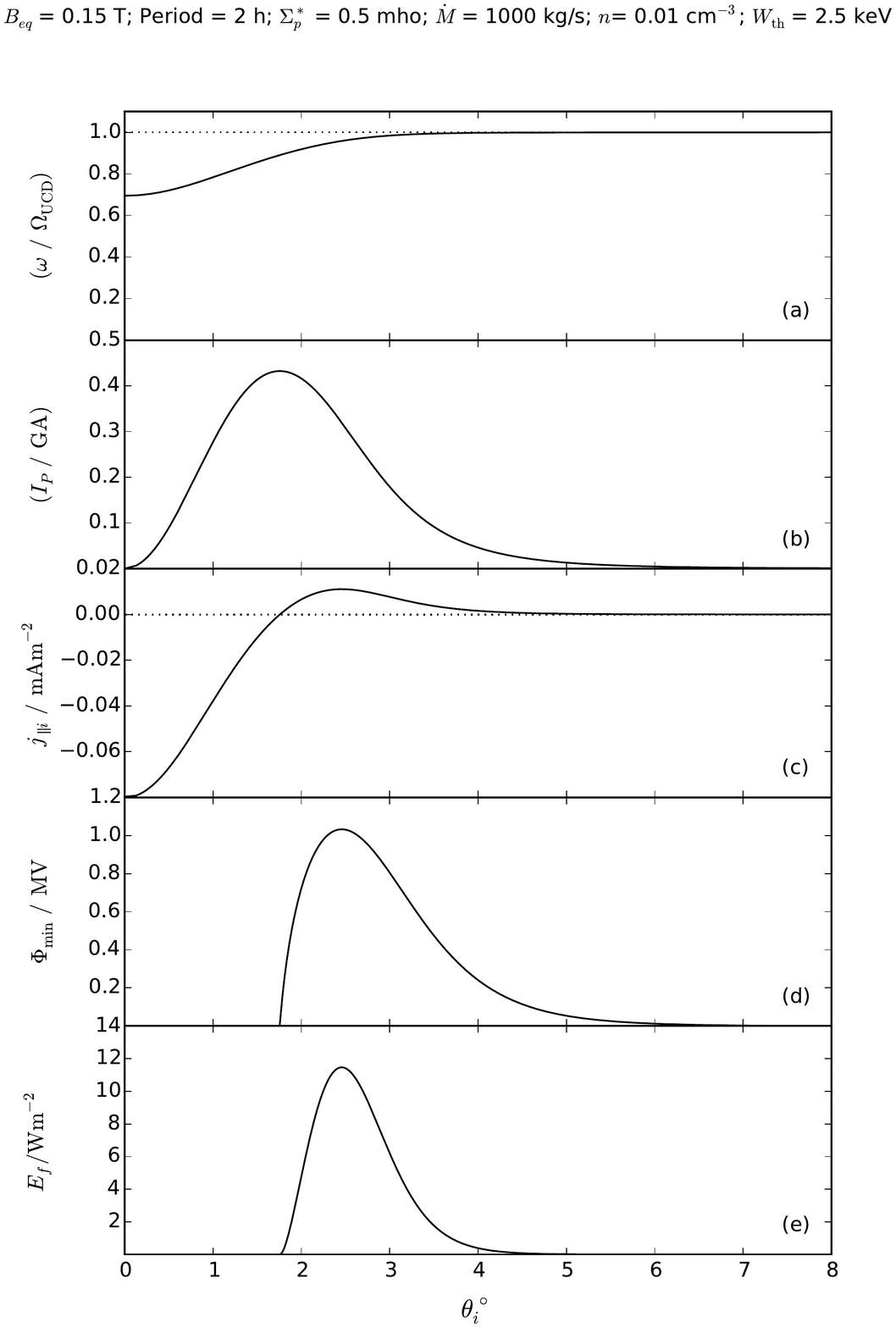}
    \caption{Counterpart profiles to those in Fig \ref{fig:equatorial}(b) - (f), now plotted versus ionospheric colatitude. Specifically we show (a) the angular velocity profile, (b) the ionospheric Pedersen current $I_P$ integrated over azimuth, (c) the field-aligned current density just above the ionosphere in mAm$^{-2}$, (d) the minimum field-aligned voltage required to drive the current shown in panel (c), and (e) the precipitating electron energy flux $E_f$. Again, the constant parameter values employed are indicated above panel (a) and are identical to those used in Fig \ref{fig:equatorial}.}
    \label{fig:ionosphere}
\end{figure}

Fig. \ref{fig:ionosphere} shows these results mapped along magnetic field lines to the ionosphere and plotted versus colatitude.  It is first evident from Fig. \ref{fig:ionosphere}(a) that the ionospheric plasma near-rigidly corotates almost everywhere, except for a small $\sim 2^{\circ}$ wide region near the pole.  Note that this contrasts with the jovian main auroral oval, which is situated at $\sim 15^{\circ}$, principally due to the much greater magnetic field strength at these objects. Fig. \ref{fig:ionosphere}(b) and Fig. \ref{fig:ionosphere}(c) highlight the narrow range of ionospheric colatitude over which the resulting currents predominantly flow.  The Pedersen current peaks at 0.43 GA at a colatitude of $\sim 1.8^{\circ}$, while the corresponding upward field-aligned current density just above the ionosphere $j_{\| i}$ peaks at $\sim 2.5^{\circ}$, dropping to negative values inside of $1.75^{\circ}$.  The field-aligned voltage $\Phi_{\mathrm{min}}$ is zero inside $1.75^{\circ}$, then rising to a peak of $\sim$1 MV at a colatitude of $\sim 2.5^{\circ}$ before dropping to zero beyond $\sim 7^{\circ}$. The precipitating electron energy flux $E_f$ peaks at 11.4 W m$^{-2}$ also at 2.5$^{\circ}$, and is zero inside 1.75$^{\circ}$ and beyond $\sim 5^{\circ}$. The resulting radio spectral luminosity obtained using equation (\ref{eq21}) and (\ref{eq22}) is 21 kW Hz$^{-1}$, which is $\sim$2 orders of magnitude below that typically observed. 

\begin{figure}
	\includegraphics[width=\columnwidth, trim=0 0 0 100]{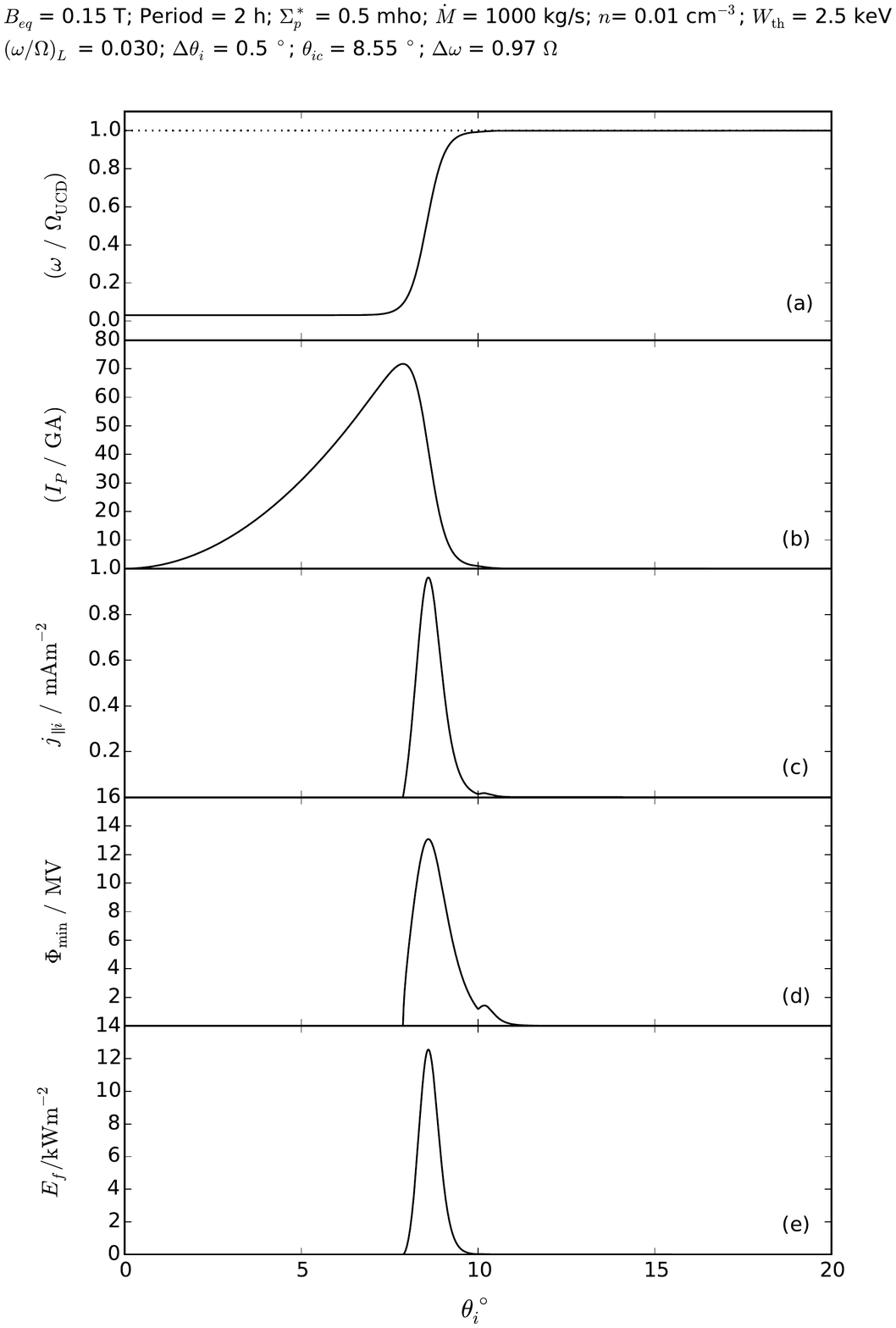}
    \caption{Ionospheric profiles equivalent to those shown in Fig. \ref{fig:ionosphere} but for the open magnetic field model.  The OCFB is fixed at 10 $^\circ$ colatitude.}
    \label{fig:open}
\end{figure}

We thus now turn to results obtained for an open magnetosphere shown in Fig. \ref{fig:open}. We have fixed the OCFB at 10$^{\circ}$ in conformity with observations at the planets in the solar system.  In this case the angular velocity of open field lines at the pole is 0.03 $\Omega_{\mathrm{UCD}}$, to which the angular velocity transitions over the OCFB from the last closed value given by the H-P equation as discussed above. In comparison with the results of Fig. \ref{fig:ionosphere} for the closed field, Figs. \ref{fig:open}(a) and (b) show that including a transition of angular velocity to the \citet{isbell1984} value at the OCFB leads to a much greater Pedersen current flowing in the ionosphere, now reaching a peak of $\sim$ 80 GA.  Consequently, the field-aligned current density (Fig. \ref{fig:open}(c)) is also enhanced such that the upward flowing portion reaches a maximum of 1 mA m$^{-2}$. The field-aligned voltage $\Phi_{\mathrm{min}}$ (Fig. \ref{fig:open}(d)) and precipitating electron energy flux $E_f$ (Fig. \ref{fig:open}(e)) are increased compared with the closed field model to $\sim$ 13 MV and 12 kW m$^{-2}$ respectively, with a resulting spectral luminosity of 0.88 MW Hz$^{-1}$, closer to observation.

\subsection{Parameter space investigation}

In the above section we demonstrated the model using approximate jovian values for the M-I coupling parameters and the thermal energy and number density of the source plasma population.  In the case of UCDs, the only relevant quantities known from observation are the rotation rate, the magnetic field strength, and the spectral luminosity.  The values of the other parameters present in our model are unknown and we therefore consider a range of values, determining the domains over which they are consistent with the observed spectral luminosities.

\begin{figure}
	\includegraphics[width=\columnwidth]{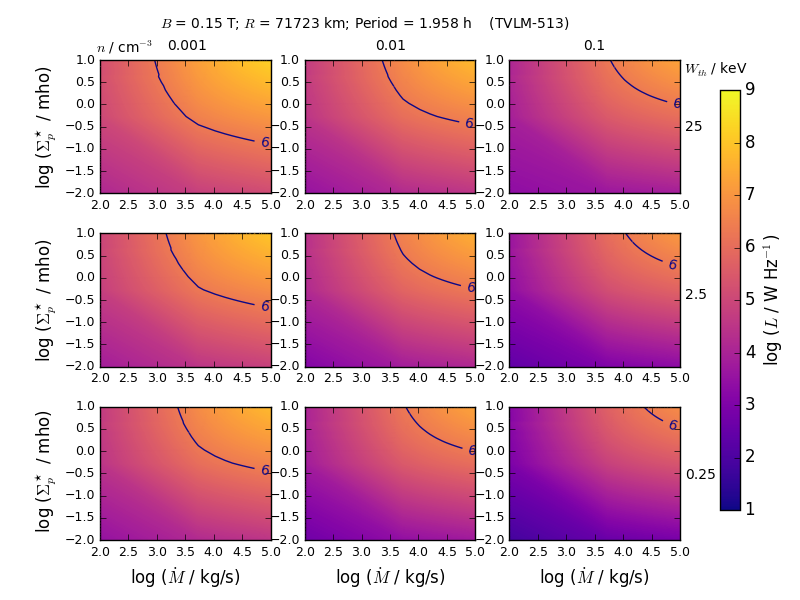}
    \caption{Contour plots of the spectral luminosity in W Hz$^{-1}$ in a simulation of the UCD TVLM-513 assuming a closed magnetosphere. Each individual panel shows the luminosity for a range of mass outflow rate $\dot{M}$ and Pedersen conductivity $\Sigma_P^*$.  The panels are arranged such that plasma population number density $n$ ranges between 0.001 and 0.1 cm$^{-3}$ and the thermal energy $W_{th}$ ranges between 0.25 and 25 keV.  The solid black line in each panel indicates a luminosity of 1 MW Hz$^{-1}$, similar to the value measured at TVLM-513.}
    \label{fig:contour}
\end{figure}

\begin{figure}
	\includegraphics[width=\columnwidth]{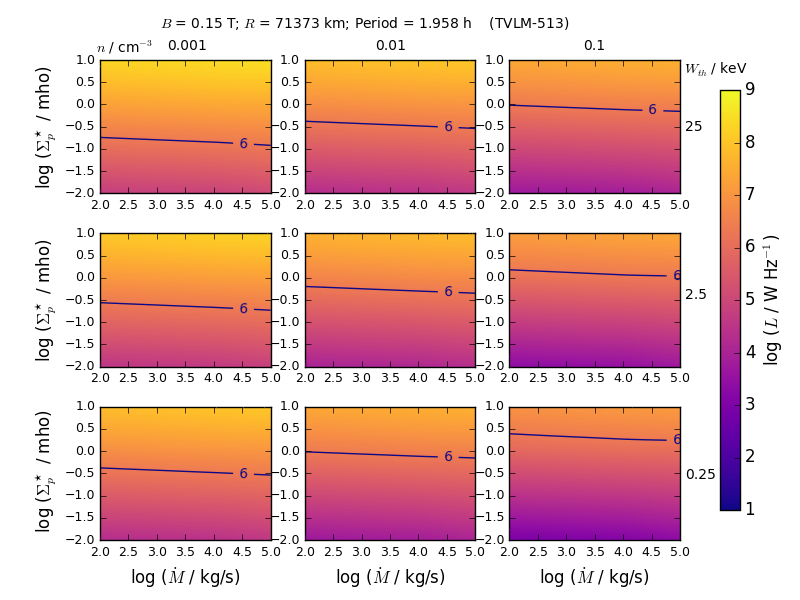}
    \caption{Equivalent plots to those in Fig. \ref{fig:contour} of the spectral luminosity of TVLM-513, in this instance assuming an open magnetic field topology with the OCFB set at $10^{\circ}$ colatitude.}
    \label{fig:opencontour}
\end{figure} 

Fig. \ref{fig:contour} shows a set of plots of radio luminosity for TVLM-513, for the closed magnetosphere model, arranged such that the electron source population number density $n_e$ varies over the columns and the thermal energy of the source plasma $W_{th}$ varies over the rows.  Each individual panel then shows results over a wide but plausible range of the mass outflow rate and effective Pedersen conductivity, which are centred approximately around canonical jovian values. The solid black line in each panel indicates a value of 1 MW Hz$^{-1}$, i.e. the order of magnitude of the luminosity observed at TVLM-513 and other fast-rotating UCDs. The central panel uses $n =$ 0.01 cm$^{-3}$ and $W_{th} = 2.5 $ keV; values measured outside the current sheet at Jupiter by Voyager \citep{scudder1981}.  Low density values are explored since, at Jupiter, the centrifugal force acts to confine cold plasma to the equatorial region, leaving hot rarefied plasma at high latitudes in the region of ECMI generation. As discussed by N12 it seems likely that cold plasma in rapidly rotating UCD magnetospheres is similarly centrifugally confined.

We find that the closed field model is able to generate spectral luminosities of the order of magnitude as those observed at TVLM-513 and other fast rotating UCDs, although we see from Fig. \ref{fig:contour} that a significant portion of the parameter space does not produce luminosities above 1 MW Hz$^{-1}$.  The detectable regime in each plot lies in the upper right region with high $\Sigma_P^*$ and high $\dot{M}$ such that a high Pedersen conductivity and high mass outflow rate are required to produce observable luminosities in this closed magnetosphere topology. Physically the luminosity increases with $\Sigma_P^*$ as higher conductivity leads to larger current magnitudes, and hence higher radio intensity.  The luminosity decrease with decreasing mass outflow rate is due to a combination of two factors. Firstly, a smaller mass outflow rate causes breakdown of corotation to occur at larger radial distances, which means that the angular velocity gradient maps to higher latitudes in the ionosphere.  Hence, the area of the annulus which forms the auroral emission region is smaller for lower values of $\dot{M}$. Secondly, as the Hill distance increases with falling mass outflow rate the angular velocity gradient moves increasingly beyond the confines of the magnetosphere, and hence a reduction in the strength of the current system and luminosity. 

Fig. \ref{fig:opencontour} shows equivalent plots to Fig. \ref{fig:contour} but for the open magnetic field topology.  The principal difference in the case of the open magnetosphere is that the luminosity increases more slowly with mass outflow rate.  By allowing the angular velocity to transition to the \citet{isbell1984} value at the pole, and by fixing OCFB position we have largely removed the effects mentioned above that are responsible for the relation between $\dot{M}$ and $L_r$ seen in the closed magnetosphere model. However, it is important to note that for given values of $\Sigma_P^*$ and $\dot{M}$ the power is in general higher for an open magnetosphere than for a closed magnetosphere. A summary enumeration of the free parameters employed in the models is given in Table \ref{tab:params}

\begin{table}
	\centering
	\caption{The free parameters in our models and the range of values explored for each}
		\label{tab:params}
	\begin{tabular}{l | l} 
		\hline		
		Parameter & Range explored \\
		\hline
		Pedersen conductance $\Sigma_P^*$ & 0.01 - 10 mho \\
		Mass outflow rate $\dot{M}$    &   $10^2 - 10^5$ kg s$^{-1}$ \\
		Plasma number density $n$ & 0.001 - 0.1 cm$^{-3}$\\
		Plasma thermal energy $W_{th}$  &  0.25 - 25 keV\\
		\hline   
	
	\end{tabular}
 
\end{table}

\begin{figure}
	\includegraphics[width=\columnwidth]{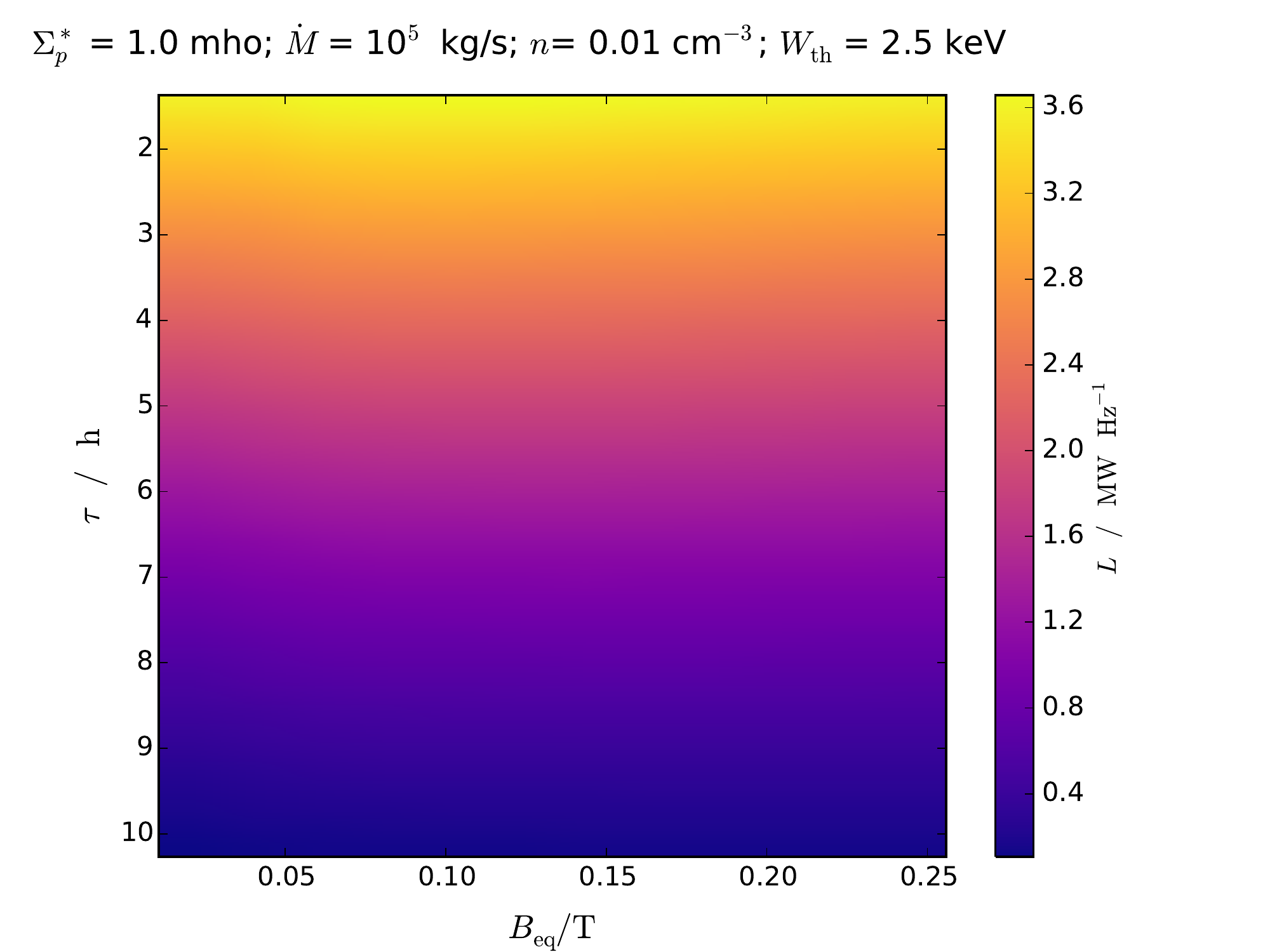}
    \caption{A contour plot for of spectral luminosity for a closed magnetosphere against equatorial magnetic field strength $B_{\text{eq}}$, and rotation period $\tau$ in hours.  Canonical jovian values are used for the plasma number density $n$ and thermal energy $W_{th}$.}
    \label{fig:contour2}
\end{figure}

\begin{figure}
	\includegraphics[width=\columnwidth]{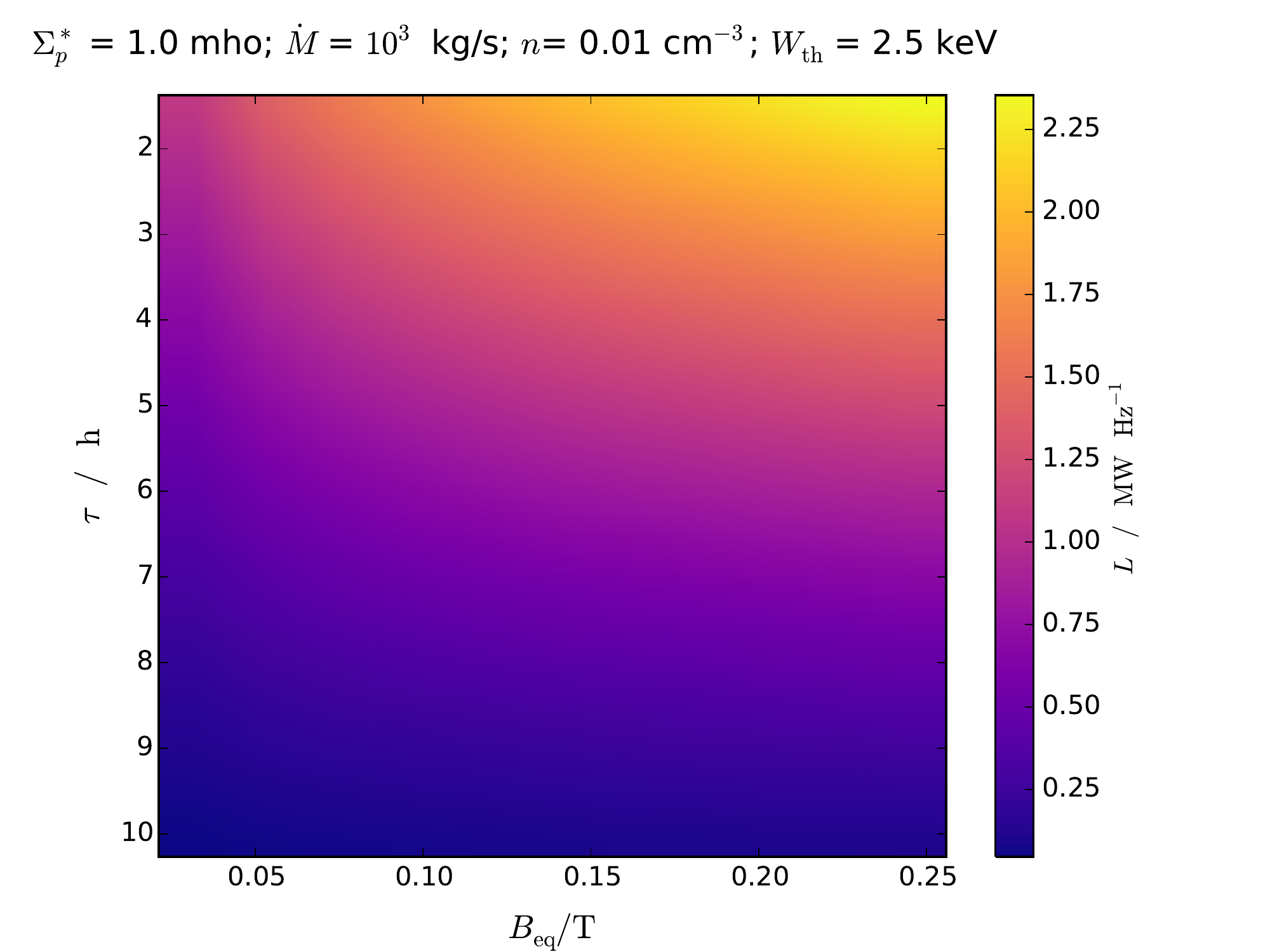}
    \caption{Spectral luminosity plot equivalent to Fig. \ref{fig:contour2} but for an open magnetic field topology.}
    \label{fig:openpower}
\end{figure}

Magnetic field strength and rotation rate are both, in principle, measurable quantities of UCDs, thus we can examine the effect of these on the spectral luminosity and compare the results with observed UCD values. Figs. \ref{fig:contour2} and \ref{fig:openpower} show the results for a closed and open magnetic field respectively, where rotation periods from $\sim$2 - 10 h, and equatorial magnetic field strengths ranging from 0.01 - 0.25 T (i.e. 25 - 600 B$_{\mathrm{Jup}}$) are plotted. Again, canonical values of the source plasma population parameters were employed along with a fiducial Pedersen conductance of 1 mho, and mass outflow rates were selected such that the luminosities obtained are typical of those that are observed. We observe that in both cases the dominant variation in the resulting luminosity is with the rotation rate of the dwarf.  The principal difference between the two cases, besides the large disparity in the magnitudes of the mass outflow rate required to generate typically observed luminosities, is the opposing relations between magnetic field strength and luminosity. In the case of the open field (Fig. \ref{fig:openpower}) luminosity increases with magnetic field strength as the larger currents drive greater auroral emission.  The opposite relation, a slight decrease in luminosity with increasing field strength, seen in the upper region of fast-rotating dwarfs in a closed magnetosphere in Fig. \ref{fig:contour2} occurs due to the Hill distance increasing more rapidly with magnetic field strength than the size of the magnetosphere $R_{\text{mp}}$, as can be seen by comparing equations (\ref{eq16}) and (\ref{eq24}).  Consequently, as the field strength is increased the corotation breakdown becomes increasingly marginal within the magnetosphere and the M-I coupling current magnitudes decrease.  Fig. \ref{fig:multimodels}(d), (e) and (f) show that while this is generally true, at low values of magnetic field strength the Hill distance lies comfortably within the magnetosphere and so an initial increase in power with field strength is seen.  

A noteworthy point arises in our model concerning the polarity of the magnetic field. In common with Jupiter we have assumed a north-south field polarity, which produces auroral emission in the region of upward flowing field-aligned current.  At Jupiter the field-aligned current density is predominantly upward flowing \citep{cnb2002}, and therefore a reversal of the magnetic field polarity would produce little to no detectable auroral emission. This motivated N12 to suggest that only 50\% of ECMI-driven radio-active UCDs may be detectable.  However, our model indicates (Fig. \ref{fig:ionosphere}(c)) that the magnitude of the negative (i.e. downward-flowing) field-aligned current density at the polar region exceeds the upward-flowing current density by a factor of approximately three.  Thus, reversing the polarity results in a beam of radio emission three times greater, and the geometry of the emitting region changes from an auroral oval to a spot centred on the magnetic pole. The same effect would be seen in the case of the open field model, but with the spot more being more broadly distributed over lower latitudes.  It is unclear, however, whether this difference in the geometry of the emitting region would be resolvable in the radio emission.

\section{Discussion} \label{sec:discussion}

In order to assess our model we now compare the results with observations.  Measurements of the spectral flux density are available for several pulsing UCDs, and we have compared our results with observations of TVLM-513 along with LSRJ1835+3259, and 2MASS J00361617+1821104 (hereafter LSR J1835 and 2M J0036), objects with well defined rotation rates and measured peaks in spectral flux density. A summary of the relevant properties of these three UCDs is given in Table \ref{tab:table1}. Follwing N12, the luminosities calculated by our model can be converted to flux density using

\begin{table*}
	\centering
	\caption{Relevant properties of the UCDs considered in this study.}%
	\label{tab:table1}
	\begin{threeparttable}
	\begin{tabular}{lccc} 
		\hline
		Property  & TVLM 513 & LSR 1835 & 2M J0036  \\
		\hline
		Polar ionospheric field strength $B_i$ (T)\textsuperscript{a}          & 0.3 & 0.3 & 0.17\\
		Rotation period (h)\textsuperscript{b}                                  & 1.96 & 2.84 & 3.08 \\
		Distance $d$ (pc)\textsuperscript{b}                                    & 10.6 & 5.7 & 8.8 \\
		Mean peak spectral flux density $F_r$ (mJy)               & 4.2\textsuperscript{c} & 2.4\textsuperscript{d}  & 0.5\textsuperscript{e}  \\
		Pulse duty cycle $\Delta \tau / \tau$                     & 0.05\textsuperscript{f}  & 0.1\textsuperscript{d}  & 0.3\textsuperscript{e}  \\
		\hline
	\end{tabular}
	\begin{tablenotes}[para,flushleft]

    \end{tablenotes}
    		\textbf{Notes:} \textsuperscript{a} From \citet{hallinan2008}; \textsuperscript{b} From Table 1 of \citet{hallinan2008}; \textsuperscript{c} From Figure 1 of \citet{hallinan2007}; \textsuperscript{d} From Figure 1 of \citet{hallinan2008}; \textsuperscript{e} From Figure 2 of \citet{hallinan2008};\textsuperscript{f} From Figure 3 of \citet{hallinan2007}. 
    \end{threeparttable} 
\end{table*}

\begin{figure*}{H}
	\includegraphics[width=\textwidth]{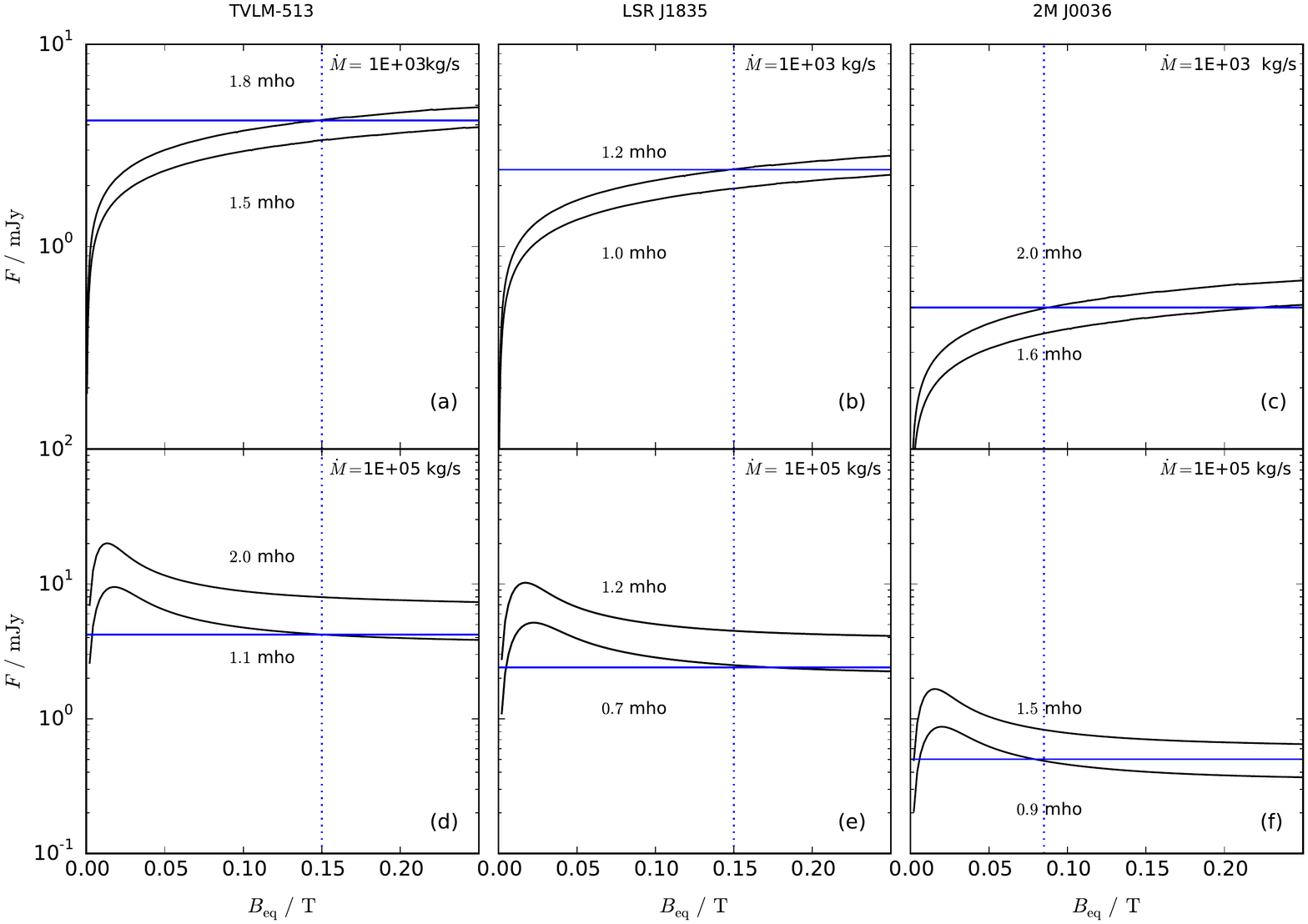}
    \caption{Plots of spectral flux density for an open magnetosphere model (top panels) and closed magnetosphere (bottom panels) for rotation rate values of three observed UCDs: TVLM-513, LSR J1835, 2M J0036. The blue horizontal line on each plot represents the approximate mean peak spectral flux density observed at the object, and the vertical line indicates the lower bound of the magnetic field strength inferred from the radio observations.  Each plot contains the results of models plotted for two separate values of the Pedersen conductance.  Values of $n = 0.01 \mathrm{\,cm}^{-3}$ and $W_{th} = 2.5$ keV are used throughout.}
    \label{fig:multimodels}
\end{figure*}

\begin{equation} \label{eq27}
	F = \frac{L_r}{\sigma d^2},
\end{equation}
where $d$ is the distance of the UCD from Earth, $F$ is the flux density, and $\sigma$ is the beamwidth. Both $d$ and $F$ are measurable quantities, and $\sigma$ can be estimated by scaling the measurements from Jupiter, for which ECMI-induced emission is beamed in 1.6 sr, giving a pulse duty cycle of 1.4 \citep{zarka2001}. Thus, on this basis we can estimate the beamwidth at a UCD by

\begin{equation} \label{eq28}
	\sigma = \left( \frac{\Delta \tau / \tau}{1.4} \right) 1.6.
\end{equation}

The rotation periods of the three dwarfs under consideration here are well constrained, though in general ECMI-induced radio observations at a specific frequency provide a lower limit to the polar field strength. Hence, here we now examine the variation of power with magnetic field strength at precise rotation rates.  In Fig. \ref{fig:multimodels}, spectral flux density $F$ is plotted against equatorial magnetic field strength $B_{eq}$ for three UCDs, in each case at the measured rotation rate of the object. The observed flux density and minimum magnetic field strength are indicated in each plot, and the results of the models described above are plotted after converting to flux density using equations (\ref{eq27}) and (\ref{eq28}). For each of the UCDs the open (top panels) and closed (bottom panels) field models are compared, with canonical jovian values taken for the source plasma population density $n$ and thermal energy $W_{th}$, and two spot values of the Pedersen conductance chosen in each plot, to indicate a plausible range that this parameter could take, given the limits on the field strengths.  The results show that in the case of an open magnetosphere (Figs. \ref{fig:multimodels}(a), (b) and (c)) a mass outflow rate of 1000 kg s$^{-1}$ is sufficient to produce spectral flux densities consistent with observations of all three UCDs at Pedersen conductances of $\sim$ 1-2 mho.  In the case of the closed magnetic field model (Figs. \ref{fig:multimodels}(d), (e) and (f)), similar Pedersen conductance values require a mass outflow rate of $10^5$ kg s$^{-1}$ to be consistent with the observed flux densities.

Since our two models produce opposing relations between magnetic field strength and luminosity, further observations of ECMI-induced emissions may be able to elucidate whether the open or closed model is the dominant mechanism, assuming the free parameters do not vary significantly between objects. We recognise, however, that this assumes all of the free parameters in the models are equal at UCDs,  emphasising the need for a significant sample of radio flux density measurements which would help to remove the effects due to Pedersen conductance, mass outflow rate, plasma number density and thermal energy. In the meantime, further ionospheric and magnetohydrodynamic modelling may shed light on the behaviour of these parameters, but this is beyond the scope of the present paper. A particularly important advance will be the advent of the modelling of the ionospheres of UCDs to estimate their Pedersen conductances, which are presently undetermined and which will further constrain these models. Knowledge of the Pedersen conductance, along with similar constraining of the other system parameters, could determine whether ECMI-induced radio emission at UCDs is driven by an outward flow of plasma from within a closed magnetosphere or by an interaction between the magnetosphere of the UCD and the ISM.

The dipole field employed in our model is unlikely to be representative of the realistic field structure at a fast rotating UCD (i.e. $\tau \sim 2-3$ h).  In common with the rotationally dominated magnetosphere of Jupiter, the field is expected to be distorted from a dipole into a magnetodisc structure by azimuthal currents arising from the requirement to balance the centrifugal force, plasma pressure, and pressure anisotropy.  This change in the magnetic field topology was shown by \citet{cnb2002} to enhance the field-aligned currents flowing at Jupiter, and narrow the width of the auroral oval and displace it to a lower latitude. An additional effect of a magnetodisc structure is to increase the size of the magnetosphere compared with the dipole, and at Jupiter the effect is a twofold increase in the size of the magnetosphere, an effect which may compensate for the effect we see in Fig. \ref{fig:contour2}, where we find only marginal breakdown of corotation within the magnetosphere. \citet{nichols2015} proposed a model to simulate the magnetodisc at Jupiter taking into account the centrifugal force, pressure and pressure anisotropy of the plasma, and future work adapting this model to a more general application at fast rotating UCDs should be undertaken to determine how this would impact the results.

As discussed earlier, enhancement of the Pedersen conductivity by precipitating electrons may occur, and thus the auroral current may itself amplify the background conductivity to levels sufficient to produce the observed radio emission.  In the absence of such modelling we have assumed a constant value throughout this paper, and ionospheric modelling of the Pedersen conductivity will indicate the magnitude and variation with field-aligned current density expected. We note that our model assumes effective magnetosphere-ionosphere coupling by Alfv\'{e}n waves, an assumption which should be examined in future work.

\section{Summary} \label{sec:summary}

We have examined the M-I coupling current system expected at UCDs and compared models for both an open and closed magnetosphere, attempting to explain the ECMI radio emission observed at some rapidly rotating dwarfs.  We considered both an ionospheric angular velocity gradient arising from a steady outward flux of angular momentum, as originally described by \citet{hill1979} and successfully applied to explain the current system generating Jupiter's main auroral oval, and also the existence of a flow shear at an OCFB.

Our results show that both of the above described processes are capable of generating luminosities commensurate with those observed at the fast-rotating UCDs considered here. Ionospheric Pedersen conductances of $\sim$ 1 - 2 mho are sufficient for a dwarf with an open magnetosphere moving relative to the ISM at 50 km s$^{-1}$, with an Io-like plasma mass outflow rate of $\sim$ 1000 kg s$^{-1}$, or for a closed magnetosphere with a steady plasma mass outflow rate of $\sim 10^5$ kg s$^{-1}$. We also provide a potential observationally testable prediction by showing that the two processes have opposing dependencies on the magnetic field strength, and thus further data may be able to discriminate between the open and closed models.


\section*{Acknowledgements}

ST was supported by an STFC Quota Studentship.  JDN was supported by an STFC Advanced Fellowship (ST/I004084/1) and STFC grant ST/K001000/1.




\bibliographystyle{mnras}
\bibliography{mybib} 



\appendix

\section{Formulation of the M-I coupling currents}

 Three distinct angular velocities with respect to a common inertial reference frame are considered, i.e. the angular velocity of the dwarf $\Omega$, the angular velocity of the neutrals in the Pedersen layer of the atmosphere $\Omega^*$, and the angular velocity of the plasma $\omega$. We expect that $\Omega^*$ will take a value reduced from $\Omega$ due to `slippage' caused by ion-neutral collisions \citep{huang1989}. Thus, for subcorotating plasma,  $\Omega > \Omega^* > \omega$, and we can write 

\begin{equation} \label{eq5}
(\Omega_{\mathrm{UCD}} - \Omega^*_\mathrm{UCD}) = k (\Omega_{\mathrm{UCD}} - \omega),
\end{equation}
where $0 < k < 1$, with the exact value unknown at UCDs.  Modelling work for Jupiter by \citet{achilleos1998} indicates a value of $k=0.5$ is reasonable, and we therefore also adopt that value here.

The height-integrated Pedersen current $i_P$ in each hemisphere is given by
 
\begin{equation} \label{eq6}
i_P = \Sigma_P E_i = \Sigma_P \rho_i  B_i (\Omega_{\mathrm{UCD}}^* - \omega),
\end{equation}
where $\Sigma_P$ is the ionospheric Pedersen conductance, $B_i$ is the ionospheric field strength in the polar region, which we assume to be radial and equal to $2B_{\mathrm{UCD}}$, and $E_i$ is the equatorward-directed ionospheric electric field strength in the rest frame of the neutral atmosphere. Combining equations (\ref{eq5}) and (\ref{eq6}) yields

\begin{equation} \label{eq7}
i_P = 2 \Sigma_P^* \rho_i B_{\text{UCD}} (\Omega_{\mathrm{UCD}} - \omega),
\end{equation}
where, in common with previous works, we have introduced the `effective' Pedersen conductance, $\Sigma_P^* = (1 - k)\Sigma_P$. Equation (\ref{eq7}) defines the Pedersen current, and, assuming north-south symmetry between the two hemispheres, it follows from the requirement of current continuity that the equatorial radial current $i_{\rho}$ is related via $\rho_e i_{\rho} = 2 \rho_i i_P$, a requirement that is also clearly evident from the topology of the current system shown in Fig. \ref{fig:schematic}.  Hence, we have 

\begin{equation} \label{eq9}
i_{\rho} = \frac{4 \Sigma_P^* \Omega_{\mathrm{UCD}} F_e}{\rho_e} \left( 1 - \frac{\omega}{\Omega_{\mathrm{UCD}}} \right).
\end{equation}
Integrating equation (\ref{eq9}) in azimuth gives the total equatorial radial current 
\begin{equation}
I_{\rho} = 2 \pi \rho_e i_{\rho} = 8 \pi \Sigma_P^* \Omega_{\mathrm{UCD}} F_e \left( 1 - \frac{\omega}{\Omega_{\mathrm{UCD}}} \right),
\end{equation}
with the corresponding azimuth-integrated Pedersen current flowing in each hemisphere simply equal to half the total radial current.

The field-aligned current can be obtained from the divergence of either the Pedersen current or the equatorial radial current.  The quantity that we require is the current density per unit magnetic field strength, which is constant along a given flux tube. Differentiation of the equatorial current, for example, gives

\begin{multline} \label{eq11}
\left(\frac{j_{\|}} {B}\right) = \frac{1}{4 \pi \rho_e |B_{ze}|} \frac{dI_{\rho}}{d\rho_e} = \\
 - 2 \Sigma_P^* \Omega_{\mathrm{UCD}} \left[ \left( \frac{F_e}{\rho_e |B_{ze}| }\right) \frac{d}{d\rho_e} \left( \frac{\omega}{\Omega_{\mathrm{UCD}}} \right) + \left( 1 -  \frac{\omega}{\Omega_{\mathrm{UCD}}}\right) \right],
\end{multline}
where we have put $|B_{ze}| = -B_{ze} $ for a field with jovian polarity. The field-aligned current density just above the ionosphere can then be found using the approximation $B_i = 2 B_{\mathrm{UCD}}$ for a dipole field, such that
\begin{equation}
j_{\| i} = 2 B_{\mathrm{UCD}} \left( \frac{j_{\|}}{B} \right),
\end{equation}
and hence 

\begin{equation} \label{eq12}
j_{\| i} = - \frac{4 B_{\mathrm{UCD}} \Omega_{\text{UCD}}}{\rho_e B_{ze}}   \frac{d}{d\rho_e} \left[ \Sigma_p^* \left( 1 - \frac{\omega}{\Omega_{\text{UCD}}} \right) F_e \right].
\end{equation}


\bsp	
\label{lastpage}
\end{document}